# Identifying Quantum Phase Transitions using Artificial Neural Networks on Experimental Data

Benno S. Rem[1,2], Niklas Käming[1], Matthias Tarnowski[1,2], Luca Asteria[1], Nick Fläschner[1], Christoph Becker[1,3], Klaus Sengstock[1,2,3,*], Christof Weitenberg[1,2].

[1]ILP – Institut für Laserphysik, Universität Hamburg, Luruper Chaussee 149, 22761 Hamburg, Germany.

[2]The Hamburg Centre for Ultrafast Imaging, Luruper Chaussee 149, 22761 Hamburg, Germany.

[3]ZOQ – Zentrum für Optische Quantentechnologien, Universität Hamburg, Luruper Chaussee 149, 22761 Hamburg, Germany.

**Machine learning techniques such as artificial neural networks are currently revolutionizing many technological areas and have also proven successful in quantum physics applications. Here we employ an artificial neural network and deep learning techniques to identify quantum phase transitions from single-shot experimental momentum-space density images of ultracold quantum gases and obtain results, which were not feasible with conventional methods. We map out the complete two-dimensional topological phase diagram of the Haldane model and provide an accurate characterization of the superfluid-to-Mott-insulator transition in an inhomogeneous Bose-Hubbard system. Our work points the way to unravel complex phase diagrams of general experimental systems, where the Hamiltonian and the order parameters might not be known.**

Ultracold quantum gases have established as a formidable experimental platform to study paradigmatic quantum many-body systems in a well-controlled environment *(1, 2)*. Important break-throughs include the realization of paradigmatic condensed matter models such as the Mott insulator transition or topological quantum matter. While these systems offer complementary observables to solid state systems, finding proper observables for quantum phases remains a key challenge, in particular in exotic systems such as non-local topological order of many-body localization. Here we explore a new approach building on modern machine learning techniques *(3)*. Inspired by the success of convolutional neural networks in image recognition, we feed such networks with single images of momentum-space density, which are a standard experimental output of quantum gas experiments. We train it on large data sets of labelled images taken far away from the phase transition and apply the trained network to test data across the phase transition. The network is able to identify the correct position of the phase transition in parameter space from single experimental images. This is crucial advance for optimizing parameters, because the phase can now be determined from single images for direct decisions in the laboratory, and points towards future fully automated quantum simulators. We expect these techniques to be valuable also for in-situ snapshots as captured by quantum gas microscopes *(4, 5)*. Similar approaches were previously applied to numerical Monte Carlo simulations of various physical models *(6-13)*. Neural networks are also opening new avenues in other areas of quantum physics, such as the representation of quantum many-body states *(14, 15)* or the optimization of complex systems *(16-18)*.

We demonstrate the power of artificial neural networks on two physical examples, namely the topological phase transition in the Haldane model and the superfluid-to-Mott-insulator transition in the Bose-Hubbard model, both realized for cold atoms in optical lattices. We show that we can perform tasks, which were not possible with conventional techniques, such as the determination of non-local topological order from a single-shot image – a problem for which there is no physical model – and a much more precise localization of the superfluid-to-Mott-insulator transition compared to the usual determination via interference contrast. Our examples show that machine learning can help in the identification of observables that were not previously obvious.

In a first set of experiments, we consider the Haldane model on the honeycomb lattice *(19)*, a paradigmatic model for topological bands with possible Chern numbers $C = -1, 0, 1$ without a net magnetic flux (inset in Fig. 2B). The topological phase diagram is spanned by the Peierls phase $\Phi$ and the sublattice energy offset $\Delta_{AB}$, which quantify the breaking of time-reversal symmetry and inversion symmetry, respectively, with the phase boundaries given by $\Delta_{AB} = \pm 3\sqrt{3} J' \sin(\Phi)$ *(19)*. We realize a Haldane-like Hamiltonian as an effective Floquet Hamiltonian for ultracold fermionic atoms in a driven honeycomb optical lattice *(20-23)*. In this Floquet realization, the inversion symmetry breaking is controlled via the shaking frequency $\omega$ and the time-reversal symmetry is controlled by the shaking phase $\varphi$ *(24)*. In the experiments, we adiabatically ramp up the shaking amplitude and ramp the shaking frequency in order to populate the lowest band. After different waiting times in the Floquet system, we suddenly switch off all trapping potentials, such that the cloud expands during a certain time-of-flight and the momentum distribution is mapped onto the real-space density distribution, which we capture with an absorption image (Fig. 2A). These momentum space images are directly fed into the convolutional neural network. The principle suitability of the momentum-space density is motivated by state tomography methods, which rely on such images and allow the full measurement of the Berry curvature *(22, 24, 25)*. However, there is no physical model for how to extract the Chern number from single images and thus the Chern number is not obvious for the human eye *(22)*.

We take data for varying shaking frequencies and shaking phases, thus mapping out the topological phase diagram. For training the network, we use data far away from the phase transition, where we can unambiguously label them with a Chern number $C = -1$, 0, or +1. We then use the trained network to classify the images in the transition region. The network outputs the probabilities for each class $P_C$ (Fig. 2B). Along a cut through the phase diagram for circular shaking ($\varphi = -90°$), the network predicts a single Chern number with near unit certainty away from the phase transitions, even outside the training regions. Moreover, it shows a smooth cross over between two Chern numbers in a small transition region with a full width of 100-200 Hz, which is due to the inhomogeneity of the system.

We use the same trained network to map out the entire two-dimensional phase diagram using just a few images per parameter set. In Fig. 2C we plot the expectation value of the Chern number $\langle C \rangle = \sum_{C=-1}^{1} C \cdot P_C$ as a function of the shaking frequency and shaking phase. The network identifies the two lobes with Chern numbers $-1$ and $+1$, which are characteristic for the Haldane model, in quantitative agreement with a numerical calculation of the Floquet system *(24)*. The identification from single snapshots allows mapping out the full 2D Haldane phase diagram. We note that this has been unfeasible with traditional methods. As a nice illustration of the robustness of the method, the network which was trained at a lattice depth of $V = 7.4 \, E_r$, also correctly predicts the topological phase transitions for other lattice depths (Fig. 2D). Furthermore, while experimental issues such as a significant population of the second band or the comparison of different phases within the micromotion of the Floquet system often cause difficulties in conventional data analysis *(21, 22, 26)*, the network is successfully trained to account for them, because they are present in the training data.

As a second example, we study the Bose-Hubbard model, which is a paradigmatic model for strongly-correlated physics and a pioneering model for quantum simulations with ultracold atoms *(27-29)*. The Hamiltonian describes the competition between the kinetic energy quantified by the tunnel element $J$ and the interaction energy quantified by the on-site interaction $U$ (see inset to Fig. 3B). In a cold atom realization, the parameters $J$ and $U$ can be tuned via the lattice depth $V$ *(29)*. At a commensurate filling of $n$ atoms per site, the model hosts a quantum phase transition from a superfluid phase to a Mott insulating phase. In a mean field approximation, the transition occurs at $U/zJ = 5.8; 9.9; 13.9$ for a filling of $n = 1; 2; 3$ *(27)*. Here $z$ is the number of nearest-neighboring lattice sites, which is 6 for the two-dimensional triangular lattice considered here *(30)*. Two different correlated calculations for the triangular lattice predict the transition around $U/6J = 4.43 - 4.9; 7.53 - 8.3; 10.6 - 11.9$ for $n = 1; 2; 3$ *(31, 32)*. With an external trap, different fillings are realized within the system, which can be described with a local density approximation and form a sequence of shells with commensurate densities in the Mott insulating regime, as directly observed with high-resolution in situ imaging *(4, 5, 33)*.

Here, we study the phase transition using the momentum distribution with its characteristic Bragg peaks indicating the coherence in the system (Fig. 3A) *(29, 34)*. We feed the momentum-space images into a neural network and study its identification of the phase transition (see Fig. 3B). We train the neural network with data far in the superfluid regime ($1.3 < U/6J < 2.2$) and far in the Mott insulator regime ($120 < U/6J < 415$) using a total of 557 images. When the trained network is applied to classify the data from the intermediate lattice depths, we find two clear plateaus for the superfluid phase and the Mott insulating phase, as well as a well-defined transition region for $4.3 < U/6J < 10.9$ (for $0.95 > P_{SF} > 0.05$), where the probability for the superfluid phase $P_{SF}$ decreases from 1 to 0. The transition region precisely fits with the theoretically expected parameter values, for which successive Mott shells form in the inhomogeneous system such that the fraction of particles in the superfluid phase decreases *(24)*. For our estimated particle numbers, we expect shells of $n = 1$, $n = 2$ and $n = 3$ to form. We have analyzed the robustness of these values to the choice of training regions.

For comparison, we also plot the visibility and the condensate fraction obtained from the same data, which are conventionally used to characterize the transition *(29, 30, 34, 35)*. These quantities only drop smoothly as a function of $U/6J$ with no clear indication of the location of the phase transition. While the visibility, which captures the short-range coherence of the system, remains finite deep into the Mott insulating region due to coherent particle-hole admixtures, the signal from the neural network reaches an unambiguous identification of the Mott insulating phase with $P_{SF} \approx 0$ for $U/6J > 10.9$. Our analysis shows that the characterization of the superfluid-to-Mott-insulator transition in the Bose-Hubbard model from momentum-space images does enormously profit from machine learning techniques, which yield a much smaller transition region as compared to the conventional analysis using either the visibility or the condensate fraction.

In conclusion, we have demonstrated the identification of phase transitions via machine learning techniques on common experimental momentum-space images using the example of a Chern insulator and a Mott insulator. Our results point the way to unravel complex phase diagrams of experimental systems, where the Hamiltonian and the order parameters might not be known. Future directions include unsupervised machine learning algorithms *(9, 13, 36)*, i.e. with unlabeled data, which might be useful, e.g., for the experimental identification of interacting topological systems. Machine learning approaches can also be used for reconstructing challenging many-body quantities such as entanglement entropy, which are experimentally hard to access *(37)*. Furthermore, cold atom systems are a promising platform for realizing quantum machine learning ideas, which combine machine learning techniques with the speed-up of quantum computers *(38)*.

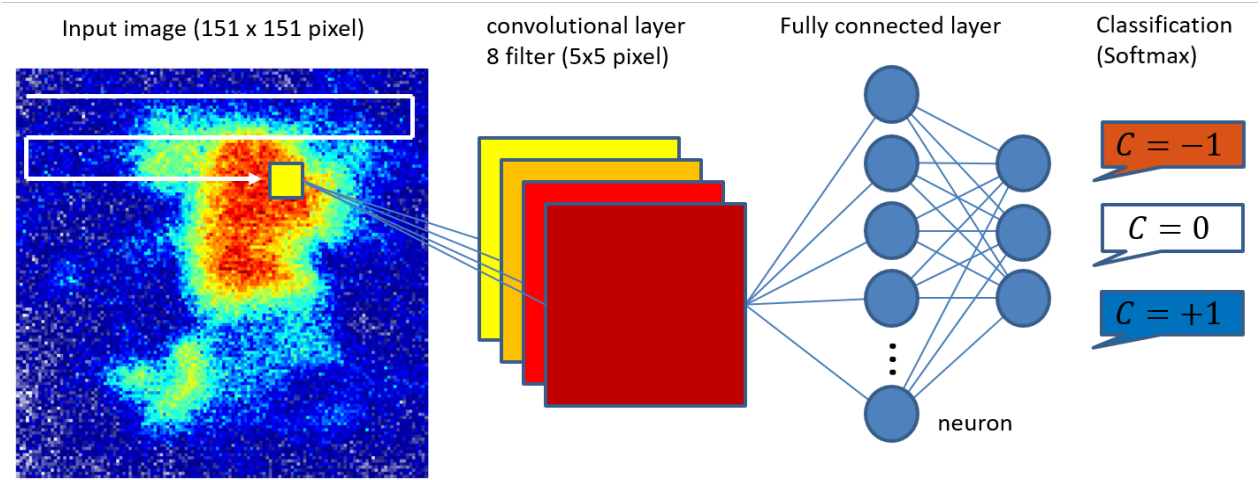

**Fig. 1. Using a neural network to identify physical phases from experimental images.** Single images of the density of atoms in momentum space after time-of-flight (false color representation of a single-channel image) serve as input for a deep convolutional neural network with a variety of layers including convolutional filters and fully connected layers. The white line represents the sliding of the filters across the input image. The final softmax layer outputs the probability that the image belongs to one of the classes (here Chern numbers $C = -1$, 0 or 1). The weights of the network are trained on many labeled images and the network can then classify an unknown single image with high confidence. This approach – originally developed for image recognition – works well for identifying physical quantum phases from experimental images.

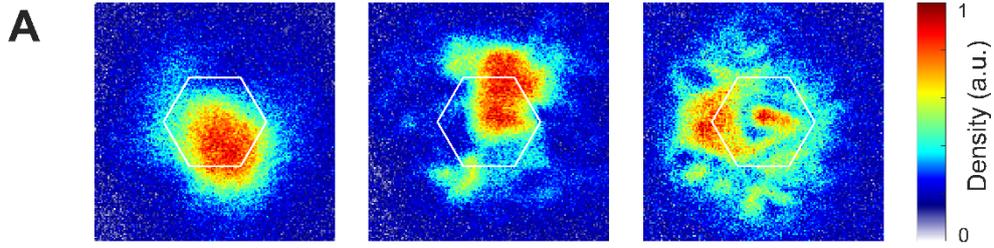
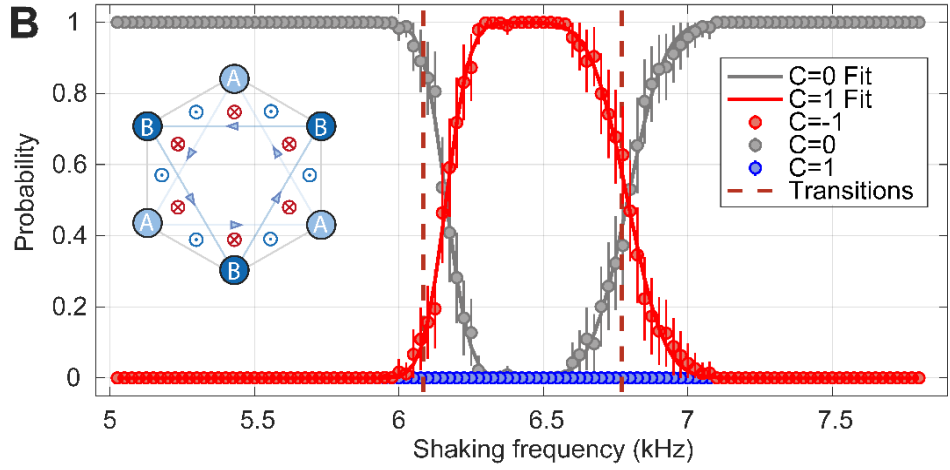
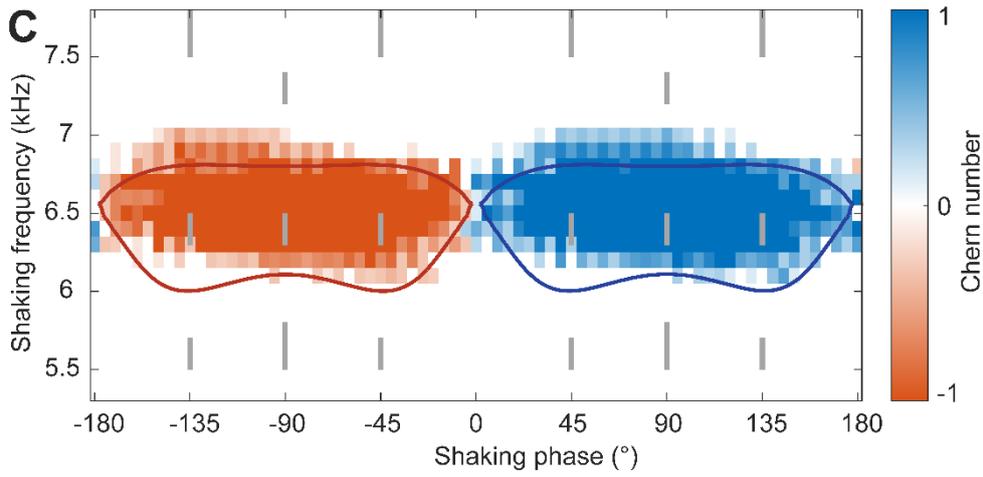
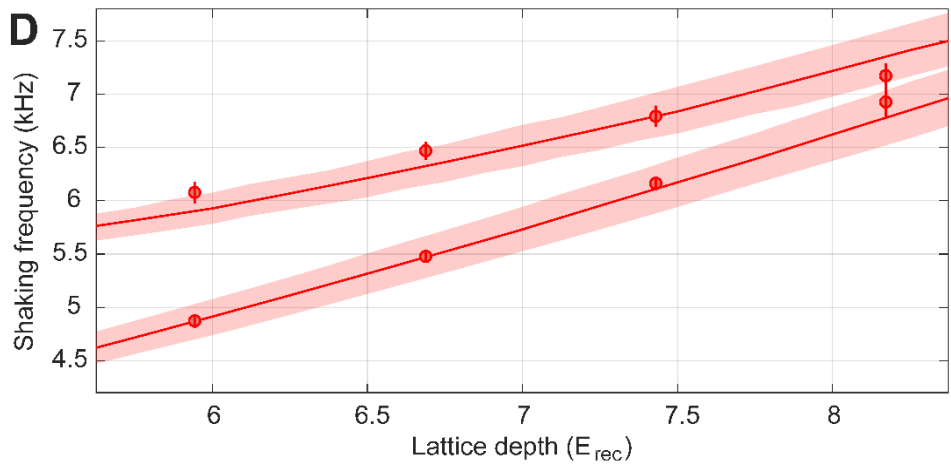

**Fig. 2. Mapping out a topological phase diagram using a neural network.** (**A**) Examples of single experimental images of ultracold fermionic atoms released from the driven optical lattice. The shaking phase is $\varphi = -90°$ and the shaking frequencies are $\omega/2\pi = 5.0$ kHz ($C = 0$), $\omega/2\pi = 6.4$ kHz ($C = 1$) and $\omega/2\pi = 7.8$ kHz ($C = 0$), respectively. The images are 151 x 151 pixel in size centered around zero momentum and include the full first Brillouin zone (white hexagon). (**B**) Probability for the different Chern number classes as identified by the trained neural network. The network was trained for Floquet frequencies far away from the phase transitions (grey, thin short lines in C). The probability is averaged over the results for 47 individual images and the error bars denote Clopper Pearson 68% confidence intervals using the Wald method. Fitting an error function to the data lets us identify the positions of the phase transitions at 6.124(3) kHz and 6.869(3) kHz by taking the 50% point of the probabilities. The dashed lines show the transitions as expected from an ab initio numerical calculation. The inset illustrates the tight-binding scheme of the Haldane model with the staggered fluxes in the subplaquettes. (**C**) Haldane-like phase diagram of the Floquet system obtained from 10,436 evaluated test images (3-7 images per parameter) using a neural network trained at the parameters indicated by the grey lines (in total 15,963 images for training and 3,992 images for validation of the network). The solid lines indicate the predicted phase transitions from our *ab initio* numerical calculation. (**D**) The circles show the positions of the phase transitions for circular shaking ($\varphi = -90°$) at varying lattice depths $V$ identified by a network trained with the data at $V = 7.4\ E_r$. The error bars denote the width of the error function fitted to the network output as in B. The lines show the predicted phase transitions from our *ab initio* numerical calculation with the grey regions denoting the systematic uncertainty calculated for an error of 0.2° on the polarization of the lattice beams.

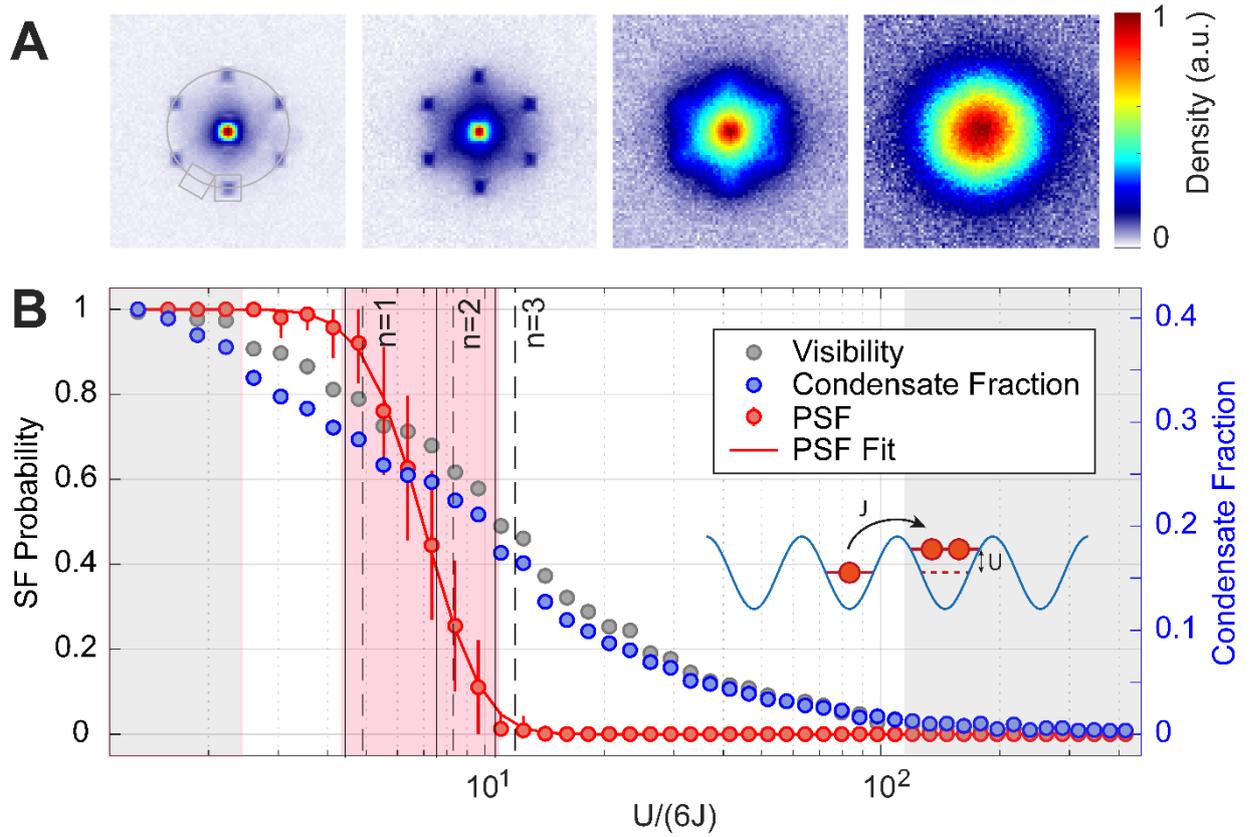

**Fig. 3. Characterizing the superfluid-to-Mott-insulator transition**. (**A**) Examples of single-shot momentum-space images of cold atoms released from a triangular optical lattice for $U/6J = 1.3, 7.3, 20.6,$ and $133$. The Bragg peaks at multiples of the reciprocal lattice vector indicate the coherence in the system for small $U/6J$. The visibility of the Bragg peaks is evaluated at the indicated boxes at constant distance from the center in order to remove the influence from the density envelope. The condensate fraction is obtained from bimodal fits to the diffraction peaks *(30)*. (**B**) Visibility, condensate fraction and the results from the trained network for the probability for the superfluid phase $P_{SF}$ as a function of $U/6J$. Visibility and condensate fraction drop smoothly, while the probability curve obtained from the network clearly defines a superfluid region, a Mott-insulating region and a transition region (red area, $4.3 < U/6J < 10.9$) in agreement with the predictions for the transition at filling $n = 1$, $n = 2$ and $n = 3$ (black solid lines *(31)* and dashed lines *(32)*), which occur in the inhomogeneous system. The network was trained with data far in away from the phase transition (grey areas). The inset illustrates the hopping term and the on-site interaction term of the Hubbard model.

We thank Maciej Lewenstein for stimulating our interest in machine learning of quantum phase transitions and Alexandre Dauphin and Joseph Thywissen for useful discussions. The computational resources were provided by the PHYSnet-Rechenzentrum of Universität Hamburg and we thank Bodo Krause-Kyora and Martin Stieben for technical support. We acknowledge financial support from the Deutsche Forschungsgemeinschaft via the Research Unit FOR 2414 and the Collaborative Research Center SFB 925. BSR acknowledges financial support from the European Commission (Marie Skłodowska Curie Fellowship ISOTOP, grant number 652837).*Correspondence to: klaus.sengstock@physnet.uni-hamburg.de

**Supplementary Material**

Experimental setup and system preparation. The experiments are performed with ultracold atoms in optical lattices. For the Haldane model, we use spin-polarized fermionic $^{40}$K atoms (mass $m = 40$ a.m.u.) in a honeycomb optical lattice formed by three lattice beams *(22, 30, 39)* of wavelength $\lambda = 1064$ nm intersecting under an angle of 120°. By controlling the polarization of the lattice beams, we produce an energy offset between the A and B sublattices of $h \cdot 6.1$ kHz for the lattice depth of 7.4 $E_r$ in units of the recoil energy $E_r = h^2/2m\lambda^2$ *(22)*. The external trapping frequency of about 70 Hz is dominated by the finite size of the lattice beams. We shake the lattice by modulating the phases of the three interfering laser beams with $\delta v_{2,3} = 2\Delta v[\pm \cos(\omega t + \varphi) + \sqrt{3} \sin(\omega t)]$ and $\delta v_1 = 0$, respectively. Here, $\omega/2\pi$ is the shaking frequency, $\Delta v$ is the shaking amplitude, and $\varphi$ is the shaking phase. The formula describes an elliptical forcing, which produces a Floquet system with new properties *(39, 40)*. For $\varphi = 0, 180°$, the shaking is linear (under $\pm 45°$ with respect to a reciprocal lattice vector) thus preserving time-reversal-symmetry and for $\varphi = \pm 90°$, the shaking is circular of either chirality and maximally breaks time-reversal symmetry.

For the preparation of the Floquet bands, we start with a completely filled lowest band of the static lattice and perform ramps of the Floquet parameters *(41)*. We first ramp up the shaking amplitude within 5 ms to 1 kHz at a shaking frequency of 4.5 kHz (far from the $C = \pm 1$ regimes) and then ramp the shaking frequency to the final value within 2 ms. Due to the closing of the band gaps and the heating rates, the preparation cannot be fully adiabatic and leads to a population in the lowest band of typically 50-75%. We measure the population in the lowest band by ramping back onto the bare bands, performing adiabatic band mapping and counting the population in the first and second Brillouin zones (compare Ref. *(41)*). Fig. S1 shows the fractional population in the lowest Floquet Bloch band for the different Floquet parameters as obtained by this procedure. As there is additional heating during the ramping back to the bare bands, the measurement gives a lower bound on the fractional population that is relevant to the measurements of the momentum density. The preparation works best for circular shaking ($\varphi = \pm 90°$) with a population above 70%. We attribute this to the fact that the band gap between the two Floquet bands is largest for circular shaking. The smaller population of the lowest band away from circular shaking might be the origin of the slightly larger deviation between the measured and the numerical phase diagram at these shaking phases. We choose varying waiting times in the Floquet system in order to ensure a sampling of the Floquet micromotion of the system. We image the cloud after an expansion of 21 ms time-of-flight.

For the Bose-Hubbard model, we use $^{87}$Rb atoms (mass $m = 87$ a.m.u.) in a triangular optical lattice of wavelength $\lambda = 830$ nm as described in Ref. *(30)*. For the preparation of the two-dimensional Bose-Hubbard system, we first ramp up the transversal lattice to a fixed depth of 30 $E_r$, thus creating a stack of decoupled two-dimensional systems. We then adiabatically ramp up the triangular lattice within 150 ms to prepare the system. The momentum space density is imaged after a time-of-flight of 21 ms. The Hubbard parameters $J$ and $U$ are obtained from a band structure calculation of the optical lattice potential *(2)*. The inhomogeneous system has a mean harmonic trapping frequency of $\omega_0/2\pi = 90$ Hz. The determination of the visibility and the condensate fraction is described in Ref. *(30)*.

Mean-field model for the superfluid fraction. We study a simple two-dimensional model to predict the fraction of the system in the superfluid phase as a function of $U/6J$. We use the mean-field model at zero temperature and in local density approximation and employ the experimental parameters stated above. The parameter $U/6J$ is controlled via the lattice depth as in the experiment, i.e. the interaction strength increases slightly (between $h \cdot 0.62$ kHz and

$h \cdot 1.41$ kHz) while the tunneling element decreases exponentially (from $h \cdot 57$ Hz to $h \cdot 0.48$ Hz). For each value of $U/6J$ and of the chemical potential $\mu_0$ in the center of the trap, we predict the density profile of a single two-dimensional plane. Using the local density approximation in the harmonically trapped system $\mu(r) = \mu_0 - \frac{1}{2}m\omega_0^2 r^2$ and extracting the local densities from the mean field phase diagram, we obtain the well-known shell structure with alternating superfluid and Mott insulating phases. We choose $\mu_0$ for each value of $U/6J$ to yield a constant atom number of 2000 or 4000. From this shell structure, we calculate the fraction of the atoms that are in the superfluid phase. This fraction is supposed to contribute to the condensate fraction measured in an experiment. It is plotted as a function of $U/6J$ in Fig. S2 and it decreases from 1 to 0 when the successive Mott insulating shells form. Even when all three Mott insulating shells are formed, the model predicts a significant superfluid fraction in the system. In practice, these small superfluid shells are expected to turn thermal for experimentally-attainable temperatures, because they will accommodate most entropy of the system *(2)*. The curve of the superfluid fraction gives an indication of what a proper characterization of the phase transition in the inhomogeneous system might find and, indeed, the probability for the superfluid phase $P_{SF}$ as classified by the trained network follows a similar shape with the onset of the decrease at the predicted $n = 1$ transition (compare Fig. 3).

Network architecture and implementation. Motivated by their success in image recognition, we employ convolutional layers to identify physical phases by feature maps. See, e.g., Refs. *(42-44)* for an introduction to machine learning techniques from a physics perspective. We have studied a variety of network architectures, and found the results to be quite robust to the choice of architecture. For the data presented in the main text, we used the architectures defined in Tables S1 and S2. The code is implemented in Matlab 2017b using the build-in functionality for neuronal networks and is run on GPUs. We use the technique of stochastic gradient descent with momentum implemented in the trainNetwork function in Matlab to optimize our models.

To train the network on the data of the Haldane model, we use an initial learning rate of 0.001 and we choose a mini batch size of 265. The number of training epochs is limited to 100 or until the desired accuracy on the validation data is reached. As validation data we randomly select 20% of the training images and separate them from the training images. Every 50th iteration the loss function is calculated for the validation images. We set the validation patience - which is the number of validations that the loss on the validation set can be larger than or equal to the previously smallest loss before network training stops - to 6. During training, the images are shuffled every epoch in order to avoid learning sequences of data. All other training options are the default values specified by Matlab. We do not use data augmentation for training (such as shifting, flipping, rotating or scaling of images or adding noise), because flipping the images may change physical properties of the system (see discussion below). We ensure that the amount of training data is symmetric, e.g., it contains the same number of images for $\varphi = -90°$ and for $\varphi = +90°$.

To train the networks on the Bose-Hubbard data we choose an initial learning rate of 0.001. In order to improve the learning rates, the mini batch size is set to 90. As the dataset size for the Bose-Hubbard system is significantly smaller than that for the Haldane system, we decided to not split validation data and limit the training to 50 epochs, which achieves good results for our parameters. The data are shuffled every epoch and the other parameters are the default values.

To extract the transition point from the probability curve of the classes, we fit a heuristic error function of the form and take the $P = 0.5$ value as transition point. In the case of the Haldane system, we use the form $P(x) = \frac{1}{2}\left[\text{erf}\left(\frac{x-x_1}{\sqrt{2}\sigma_1}\right) - \text{erf}\left(\frac{x-x_2}{\sqrt{2}\sigma_2}\right)\right]$ with the two transitions at $x_1$ and

$x_2$ and the widths $\sigma_1$ and $\sigma_2$. In the case of the Bose-Hubbard system, we fit the error function as a function of log(U/6J), which is approximately linearly related to the lattice depth.

Comparison to principal component analysis. The machine learning approach consists of learning the phases from images far from the transition. A similar approach could be realized by performing a principal component analysis (PCA) of the training data. Fig. S3 shows this analysis for the data of the Bose-Hubbard model. In a simplified form of PCA, we average images from the superfluid and the Mott insulating region to obtain two basis images (same ranges as the training regions for the machine learning stated in the main text). All images are normalized to the same particle number before processing. We combine these images into a matrix $b$ with dimension $2 \times N$, where $N = 87 \times 87$ is the number of pixels of a single image, and obtain the matrix $V = bb^T$, where $b^T$ is the transposed matrix. We also combine all M images to be studied in a matrix $I$ with dimension $N \times M$. By solving the linear equation $Vc = bI$ for $c$, we obtain the coefficients shown in Fig. S2, which quantify the overlap of the images with the basis images. In short, each images is decomposed as a superposition of the two basis images with the coefficients $c$. Such an analysis yields a much broader transition region and has no plateau in the superfluid region far away from the phase transition. This shows that the machine learning approach is more powerful than a simple PCA analysis and cannot be reduced to the latter. We note that a PCA analysis can be the starting point for unsupervised machine learning of phase transitions *(45, 46)*.

Discussion of network performance. We have carried out several cross checks to gain some understanding of the neural network *(47,48)*. Interpreting deep networks is still in its infancy and the networks often appear like a black box. We compared the networks of the two examples of the Bose-Hubbard system and the Haldane system and found that applying the network trained on one example to the data from the other example gives a definite result: the Haldane data are all identified as a Mott insulator instead of a superfluid and the Bose-Hubbard data are all identified as topologically trivial ($C = 0$). Identifying the data from the Bose-Hubbard system as topologically trivial is a correct classification and might point a way to using the networks for transfer learning.

In order to understand how the classification depends on the size of the image input, we repeat the training and analysis for cropped images and monitor the success probability for the validation data. In the case of the Haldane model, the training is successful down to cropping the input images to slightly less than one reciprocal lattice vector (see Fig. S4). This might be surprising, because the images do then not contain the full Brillouin zone, which is in general required to define the Chern number from a momentum-resolved Bloch state tomography. However, for a specific realization, such as the Haldane-like model studied here, non-local information of the Chern number might be directly encoded in certain local momenta. Also in the case of the Bose-Hubbard system, the training is insensitive to the image size over a broad range (Fig. S5).

As another cross check, we apply the trained network for the Haldane model to images, which were flipped along the horizontal axis, the vertical axis or both (Fig S6). Flipping the images accounts to flipping the momentum distribution. While the phase diagrams are somewhat distorted, the regions with non-trivial Chern number are still clearly identified. For a single flip, the Chern number inverts for the flip around the vertical axis, which transforms the K point into the K' point. For a discussion of the sign of the Chern number in such a Floquet realization of the Haldane model see also the supplementary material of Ref. *(23)*. A flipping around both axes corresponds to an inversion in momentum space, which changes the sign of the breaking of time-reversal symmetry and therefore leads to a change of sign in the Chern number.

To gain some intuition, why the neural network can identify the Chern number from the momentum space images released from the driven lattices, one should consider the Bloch state tomography introduced in Refs. *(22, 25)*. They rely on similar images, but require varying hold times in the static lattice before expansion for a full reconstruction of the state. While the Chern number can be obtained from a full state tomography by calculating the Berry curvature and integrating it over the first Brillouin zone, it is not obvious that this can be done unambiguously from the limited information of a single such image. Therefore our results pave a new way of analyzing phase transitions in quantum gas experiments. Also note that the state tomography method a priori cannot distinguish between a pseudospin state on the northern or southern hemisphere of the Bloch sphere *(25)* and this information usually requires an additional measurement of adiabatic band mapping. Fig. S7 shows a selection of typical single images for the three classes. The class $C = 0$ contains images for shaking frequency above and below the $C = \pm 1$ region and it is notable that the network successfully groups these two regions as one class. While the human eye can see some similarity in the images, it cannot extract the Chern number (including its sign), whereas our artificial neural network can extract it.

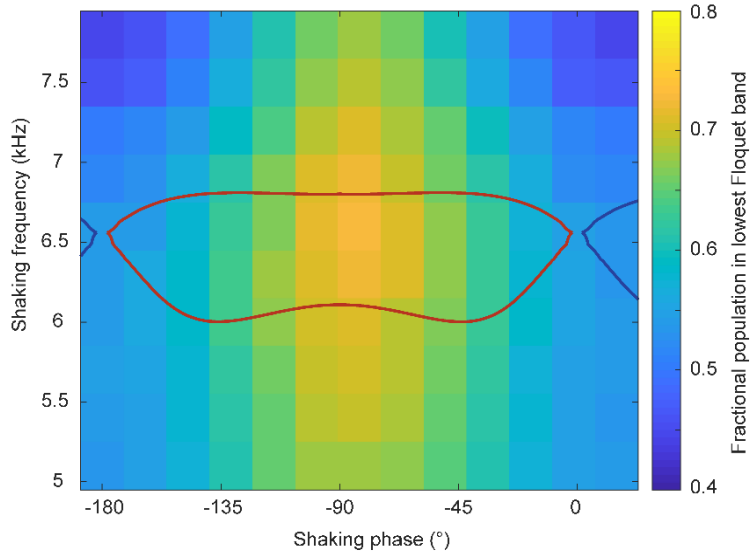

**Fig. S1. Population of the Floquet Bloch bands.** Measured fractional band population of the lowest Floquet Bloch bands as a function of shaking phase and shaking frequency. The adiabatic preparation works best for circular shaking ($\varphi = -90°$) with a population above 70%.

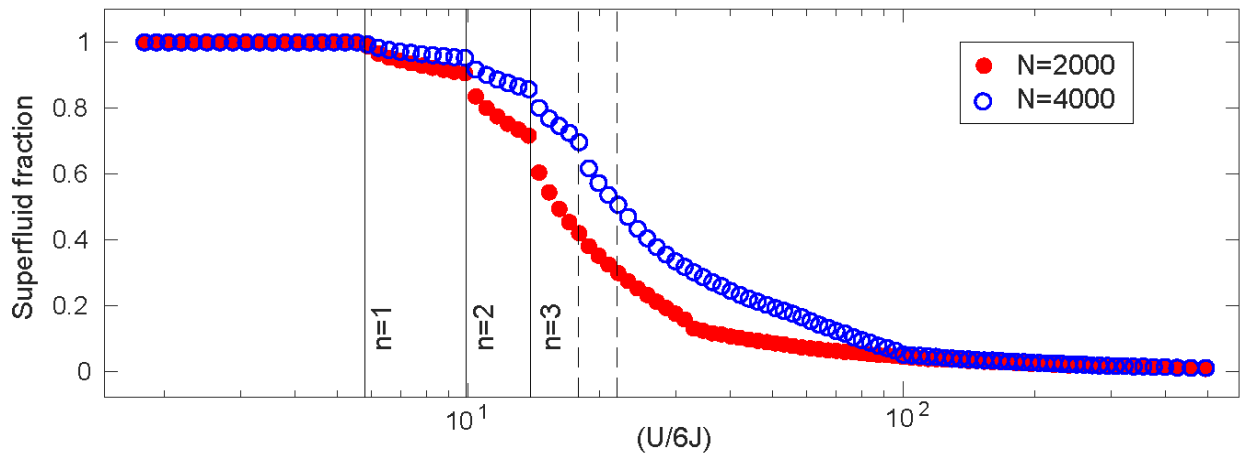

**Fig. S2. Mean-field model for the superfluid fraction.** Fraction of the atoms that are in the superfluid phase as a function of the parameter $U/6J$. The fraction begins to drop, when the first Mott shell is formed. The vertical lines show the mean-field predictions for the phase transitions of the successive Mott shells relevant to this calculation.

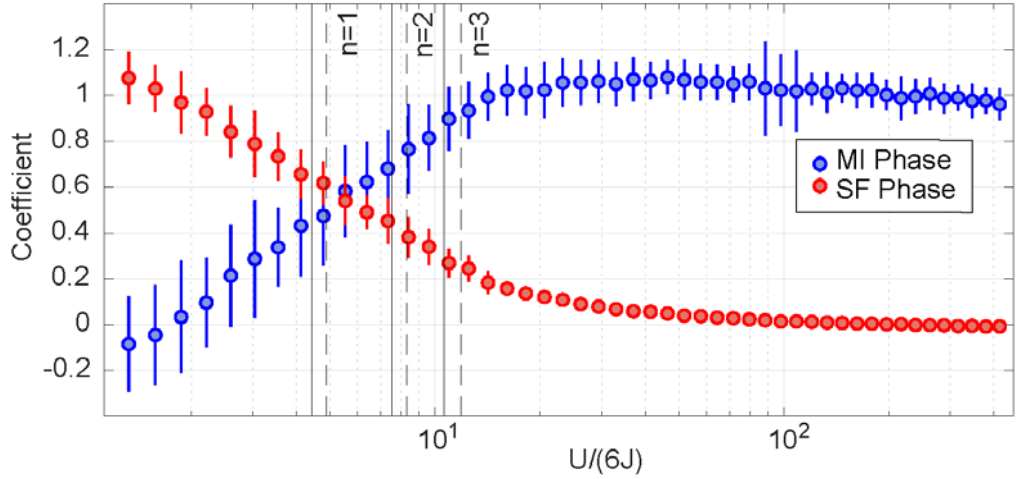

**Fig. S3. Simplified principal component analysis.** The coefficients for the superfluid phase and the Mott insulating phase quantify the overlap of the images with the respective basis images obtained by averaging many images far in the respective phases. The coefficients vary smoothly across the phase transition controlled via $U/6J$.

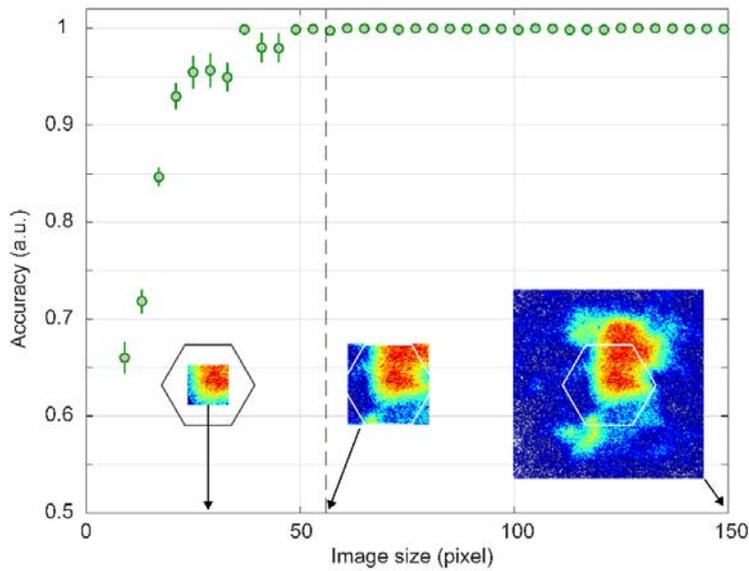

**Fig. S4. Network performance for cropped input images.** The success probability of the identification of test images from the Haldane model as a function of the image size. All images are cropped to various diameters around the center before training and evaluation. The accuracy of the identification starts dropping at around 50 pixel, which is about half the length of a reciprocal lattice vector (56 pixel, dashed line). The network and the parameters are identical to the one in the main text and the error bars stem from 5 repetitions of the training. In insets illustrate three selected cropped images.

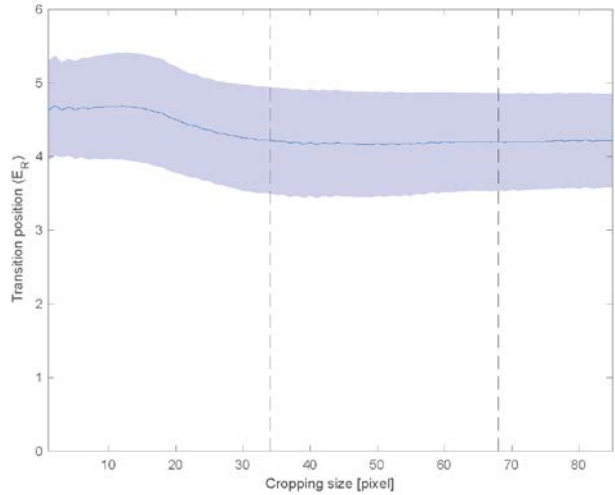

**Fig. S5. Dependence on cropped image size for the Bose-Hubbard system.** Identified position of the superfluid-to-Mott-insulator phase transition for different cropping sizes of the input images to the network. The shaded area denotes the region, where the probability for the superfluid phase drops from 95% to 5%. The dashed lines indicate the size of the reciprocal lattice vector, at which the Bragg peaks appear. The identified transition is insensitive to the image size down to around 30 px, i.e. much smaller than the reciprocal lattice vector length. This illustrated that the network does not use the visibility for the identification of the phase transition.

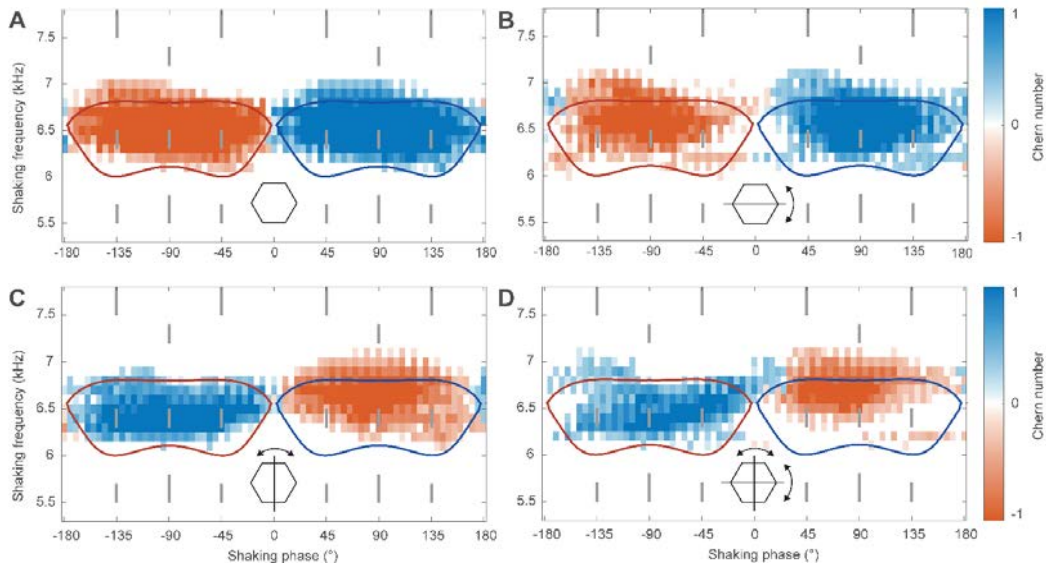

**Fig. S6. Evaluation of flipped images.** Result of the trained network applied to the evaluation images after flipping around the horizontal axis (B), the vertical axis (C) and around both axes (D). For completeness, the result without flipping is also shown (A). For C and D the sign of the Chern number is inverted.

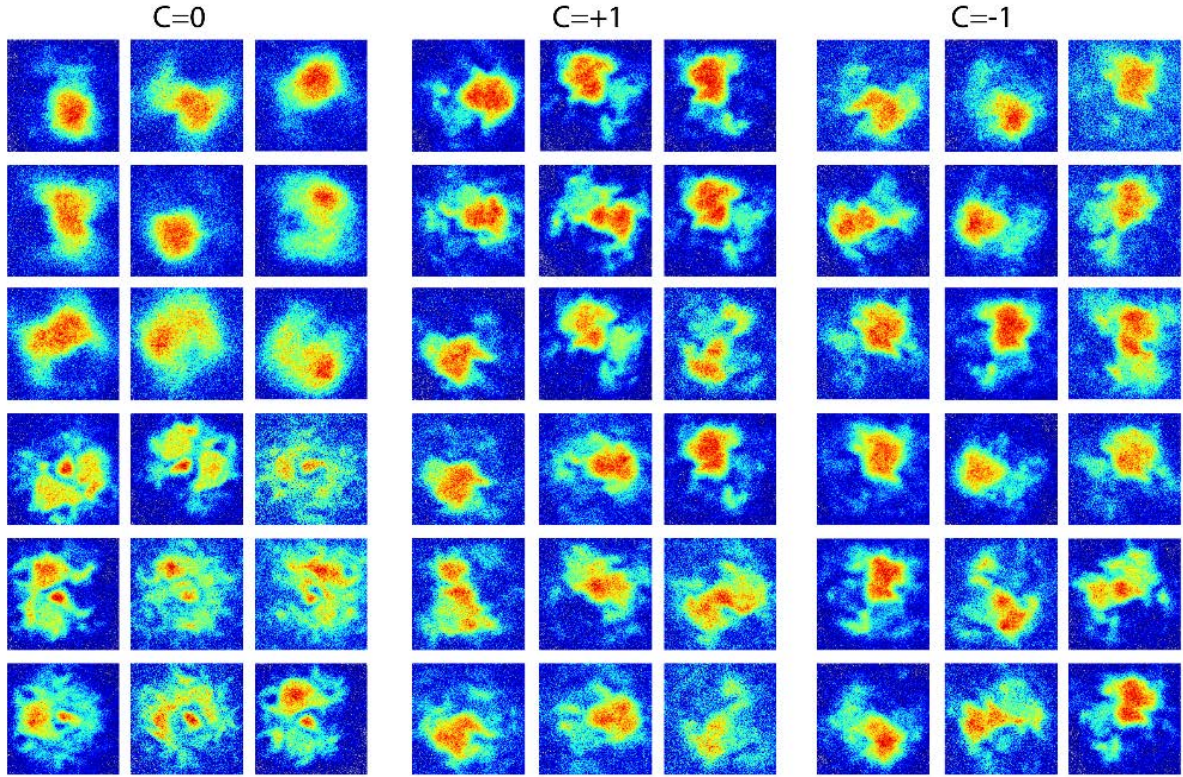

**Fig. S7. Example raw images.** Single-shot momentum space images of ultracold fermionic atoms released from the shaken optical lattice (compare Fig. 2). The network is trained with many such labeled images. The training data set has 10040 images for $C = 0$, 2859 images for $C = +1$, and 3064 images for $C = -1$; the validation data set has 2511 images for $C = 0$, 715 images for $C = +1$, and 766 images for $C = -1$. The center of mass of the atomic density varies between the images, because they sample different instances of the micromotion of the Floquet system. The Chern number and in particular its sign is not obvious to the human eye and there is no physical model for extracting the Chern number from single images. Still the artificial neural network is able to infer the Chern number of previously unseen single image.

| # | Name | Type | Description |
|---|---|---|---|
| 1 | 'imageinput' | Image Input | 151x151x1 images with 'zerocenter' normalization |
| 2 | 'conv' | Convolution | 8 5x5x1 convolutions with stride [2 2] and padding [0 0 0 0] |
| 3 | 'batchnorm' | Batch Normalization | Batch normalization with 8 channels |
| 4 | 'relu_1' | ReLU | ReLU (rectified-linear unit layer) |
| 5 | 'fc_1' | Fully Connected | 30 fully connected layer |
| 6 | 'relu_2' | ReLU | ReLU |
| 7 | 'dropout_1' | Dropout | 50% dropout |
| 8 | 'fc_2' | Fully Connected | 190 fully connected layer |
| 9 | 'relu_3' | ReLU | ReLU |
| 10 | 'dropout_2' | Dropout | 50% dropout |
| 11 | 'fc_3' | Fully Connected | 3 fully connected layer |
| 12 | 'softmax' | Softmax | softmax |
| 13 | 'classoutput' | Classification Output | crossentropyex with classes 'C=0', 'C=1', and 'C=-1' |

**Table S1. Network architecture for the Haldane system.** The architecture of the convolutional neural network consists of the 13 layers described above.

| # | Name | Type | Description |
|---|---|---|---|
| 1 | 'imageinput' | Image Input | 87x87x1 images with 'zerocenter' normalization |
| 2 | 'conv_1' | Convolution | 20 2x2x1 convolutions with stride [1 1] and padding [0 0 0 0] |
| 3 | 'batchnorm' | Batch Normalization | Batch normalization with 20 channels |
| 4 | 'relu_1' | ReLU | ReLU |
| 5 | 'maxpool' | Max Pooling | 2x2 max pooling with stride [1 1] and padding [0 0 0 0] |
| 6 | 'conv_2' | Convolution | 20 2x2x20 convolutions with stride [1 1] and padding [0 0 0 0] |
| 7 | 'crossnorm' | Cross Channel Normalization | cross channel normalization with 5 channels per element |
| 8 | 'relu_2' | ReLU | ReLU |
| 9 | 'fc_1' | Fully Connected | 150 fully connected layer |
| 10 | 'relu_3' | ReLU | ReLU |
| 11 | 'dropout' | Dropout | 50% dropout |
| 12 | 'fc_2' | Fully Connected | 2 fully connected layer |
| 13 | 'softmax' | Softmax | softmax |
| 14 | 'classoutput' | Classification Output | crossentropyex with classes 'Superfluid' and 'Mott insulator' |

**Table S2. Network architecture for the Bose-Hubbard system.** The architecture of the convolutional neural network consists of the 14 layers described above.